\documentclass[aps,prb,reprint,superscriptaddress]{revtex4-2}
\usepackage[dvipdfmx]{graphicx}
\usepackage{times}
\newcommand{\eqn}[1]{
\begin{eqnarray}
	#1
\end{eqnarray}
}
\usepackage[normalem]{ulem}
\usepackage{latexsym}
\usepackage{amssymb}
\usepackage{amsmath}
\usepackage{amsfonts}
\usepackage{lipsum} 
\usepackage[vcentermath]{youngtab}
\usepackage{upgreek}
\usepackage{bm,dsfont}
\usepackage{multirow}
\usepackage{enumitem}
\usepackage[usenames, dvipsnames]{color}
\usepackage[svgnames]{xcolor}
\usepackage{hyperref}
\hypersetup{
colorlinks = true,
linkcolor = [rgb]{0.70,0.13,0.13},
citecolor = [rgb]{0.13,0.55,0.13},
urlcolor = [rgb]{0.25, 0.41, 0.88}}
\usepackage{makecell}
\usepackage{ulem}
\newcommand{\bra}[1]{{\langle #1|}}
\newcommand{\ket}[1]{{|#1 \rangle}}

\newcommand{\ii}{\mathrm{i}}

\newcommand{\tr}{\text{tr}}

\begin{document}

\title{Two-dimensional symmetry-protected topological phases and transitions in open quantum systems}


\author{Yuxuan Guo}
\affiliation{School of Physics, Peking University, Beijing 100871, China}

\author{Yuto Ashida}
\affiliation{Department of Physics, University of Tokyo, 7-3-1 Hongo, Bunkyo-ku, Tokyo 113-0033, Japan}
\affiliation{Institute for Physics of Intelligence, University of Tokyo, 7-3-1 Hongo, Tokyo 113-0033, Japan}

\begin{abstract}
We investigate the influence of local decoherence on a symmetry-protected topological (SPT) phase of the two-dimensional (2D) cluster state. Mapping the 2D cluster state under decoherence to a classical spin model, we show a topological phase transition of a $\mathbb{Z}_2^{(0)}\times\mathbb{Z}_{2}^{(1)}$ SPT phase into the trivial phase occurring at a finite decoherence strength. To characterize the phase transition, we employ three distinct diagnostic methods, namely, the relative entropy between two decohered SPT states with different topological edge states, the strange correlation function of $\mathbb{Z}_2^{(1)}$ charge, and the multipartite negativity of the mixed state on a disk. All the diagnostics can be obtained as certain thermodynamic quantities in the corresponding classical model, and the results of three diagnostic tests are consistent with each other.  Given that the 2D cluster state possesses universal computational capabilities in the context of measurement-based quantum computation, the topological phase transition found here can also be interpreted as a transition in the computational power.
\end{abstract}
\maketitle
\date{\today}

\section{Introduction}
Understanding symmetry-protected topological (SPT) phases  \cite{ Wen2012fSPT,WenZoo1610.03911,Wen2013Classifying,Gu:2014tw,Chen2014Symmetry-protected,Pollmann:2012jw,Turner2011Topological,Bi:2015qv} has been one of the major topics in condensed matter physics over the past decade, where SPT order can be characterized by nontrivial many-body entanglement. Recently, there have been growing interest in extending the notion of SPT order to open quantum systems described by mixed states \cite{fulga2014statistical,milsted2015statistical,degroot2021symmetry,ma2022average,Paolo2023FiniteTemperature,paszko2023edge,lee2022symmetry}. While previous studies have suggested that SPT phases of density matrices could still be classified by group cohomology \cite{ChongWang2023Average,degroot2021symmetry}, a universally accepted definition and indicator of mixed-state SPT order has so far remained elusive. Given that decoherence is inevitable in any experimental system, it is of particular interest to figure out whether or not SPT order can be identified in the presence of decoherence, and if so, in what sense. 
There are, in general, two types of decoherence: $\romannumeral1)$ decoherence that can be linked to a pure state with certain ancillas, such as local bit-flip and phase errors \cite{bao2023mixedstate,fan2023diagnostics} and $\romannumeral2)$ decoherence that cannot be linked to a pure state with ancillas, such as thermalization \cite{roberts2017symmetry, huang2014topological, viyuela2014two, rivas2013density}. 
The primary goal of this paper is to focus on the former and reveal the existence of mixed-state SPT phases and their topological transitions. 

From a broader perspective, an investigation of SPT phases under decoherence has also attracted interest in quantum computation. While topologically ordered states with anyons allow for topological quantum computation \cite{bombin2015fault, shtengel2002universal, kitaev2003topological, nayak2008non}, SPT states with short-range entanglement can serve as resources for measurement-based quantum computation (MBQC) \cite{ browne2003universal, briegel2001measurement, raussendorf2001one,eisert2003measurement, monroe1996quantum}. In this context, one of the notable SPT states is a two-dimensional (2D) cluster state since it possesses universal computational power and relevant to a variety of noisy intermediate-scale quantum platforms  \cite{preskill2018quantum,endo2021hybrid,arute2019quantum,kandala2019error,bharti2022noisy}, where the effect of decoherence is crucial. This motivates the following question: are there phase transitions in the 2D cluster state under decoherence, and if so, do they signify transitions in the computational capability of MBQC?

To address the above questions, we examine the influence of local decoherence on a 2D cluster state, which is an SPT state protected by the zero- and one-form symmetries. The decoherence is modeled by local bit-flip and phase errors, which occur randomly with probabilities  $p_x$ and $p_z$, respectively. On the one hand, our investigation shows that even the presence of arbitrarily weak phase error can break the SPT order. On the other hand, however, we find that a topological phase transition can occur at a nonzero bit-flip error rate, in which case the transition can be understood as the paramagnetic (PM) to ferromagnetic (FM) phase transition in the corresponding classical spin model. These results are obtained by showing the mapping between the R\'enyi entropies of the error-corrupted mixed state and the partition functions of the Ising-type models. 

To substantiate our results and characterize the topological phase transition within the original quantum problem, we encounter two primary challenges. Firstly, the transformation of the pure state into an error-corrupted mixed state can be achieved through continuous unitary evolution involving ancilla qubits. As a result, no local order parameters are expected to exhibit singular behavior. Secondly, SPT phases lack long-range entanglement, and there exist no readily available global measures to capture their entanglement structure, such as the topological entanglement entropy being used for detecting topological order.
To address these challenges, we employ the following three distinct diagnostic tools: $\romannumeral1)$ relative entropy between two decohered SPT states with different topological edge states, $\romannumeral2)$ strange correlation function \cite{you2014wave,wu2015quantum,vanhove2018mapping,zhou2022detecting,lepore2022strange,scaffidi2016wave,zhang2022strange}
 of the error-corrupted mixed state, and $\romannumeral3)$ multipartite negativity of the decohered SPT state on a disk, which is an open-system analogue of multipartite entanglement entropy \cite{zeng2015gapped,zeng2016topological,fromholz2020entanglement}. Notably, these three measures possess the unique property of being mapped onto certain observables within the corresponding classical spin model and consistently exhibit a phase transition at the same error rates. Our mapping thus provides a bridge between decohered mixed SPT states and classical statistical mechanical models, allowing us to elucidate the nature of topological phases in open quantum systems.

The remainder of the paper is organized as follows. In Sec.~\ref{Sec2}, we introduce the stabilizer Hamiltonian of the cluster state on the Lieb lattice and analyze its symmetries, ground states, and emergent boundary degrees of freedom. In Sec.~\ref{Sec3}, we investigate how local quantum channels alter the pure cluster state and provide a perspective on the condensation of domain walls within the decohered mixed state. In Sec.~\ref{Sec4}, we introduce three distinct diagnostic measures and elucidate their corresponding interpretations and phase transitions within the framework of classical spin models.  In Sec.~\ref{Sec5}, we give a summary of results and suggest several directions for future investigations.

\section{Lattice model for a 2d cluster state}
\label{Sec2}
Lieb lattice is an edge-decorated 2D square lattice, which has two inequivalent sub-lattices. Sub-lattice $A$ is defined on the vertex of a $N\times N$ square lattice and sub-lattice $B$ is defined at the edges of the square lattice. We put a spin-$\frac{1}{2}$ on each of $A$ and $B$, which is labeled by $\tau_v$ and $\sigma_e$, respectively; see Fig.~\ref{fig:main}(a). A 2D cluster state is given by the ground state of the following frustration-free stabilizer Hamiltonian:

\begin{figure}[b]
    \centering
    \begin{minipage}[t]{0.22\textwidth}
        \centering
      \includegraphics[width=\textwidth]{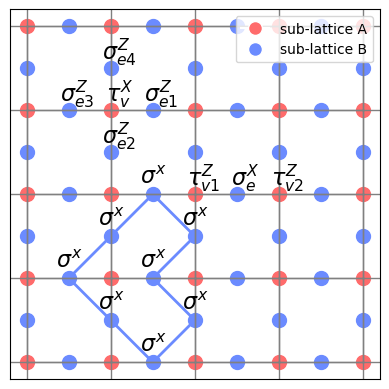}
    \end{minipage}
    \hfill
    \begin{minipage}[t]{0.22\textwidth}
        \centering
    \includegraphics[width=\textwidth]{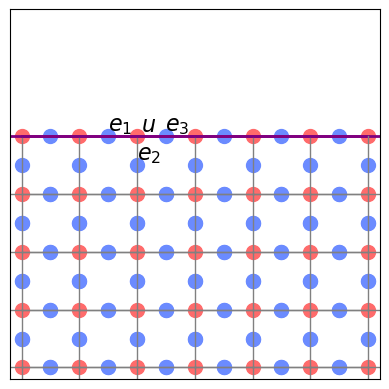}
    \end{minipage}
    \caption{(Left) Schematic illustration of the 2D cluster state. The stabilizer Hamiltonian~\eqref{stabilizer} gives rise to the cluster state (also known as a graph state). The blue line indicates a generator of $\mathbb{Z}_2^{(1)}$ symmetry. (Right) Emerging boundary spin degrees of freedom. The black lines delineate the boundary of the Lieb lattice, and a localized spin-$\frac{1}{2}$ can be defined at each vertex along the boundary.}
    \label{fig:main}
\end{figure}

\begin{gather}
\label{stabilizer}
    H_{\text{Cluster}}=-\sum_{v \in A} A_v-\sum_{e\in B}B_e, 
\end{gather}
where $A_v=\tau_v^x \sigma_{e_1}^z \sigma_{e_2}^z \sigma_{e_3}^z \sigma_{e_4}^z$ and $B_e= \sigma_e^x \tau_{v_{1}}^z \tau_{v_2}^z$. We note that all $A_v$ and $B_e$ mutually commute with each other and satisfy $A_v^2=1$, $B_e^2=1$. The ground state of $H_{\text{Cluster}}$ under the periodic boundary conditions (PBC) is uniquely characterized by the conditions $A_v=B_e=1$ for all $v,e$:

\begin{gather}
\rho_{\text{SPT}}^{\text{(PBC)}}=\prod_{v\in A}\frac{1+A_v}{2}\prod_{e\in B}\frac{1+B_e}{2}.
\end{gather}

The 2D cluster state is protected by $\mathbb{Z}_2^{(0)}\times\mathbb{Z}_2^{(1)}$ symmetry. Here, $\mathbb{Z}_2^{(0)}$ represents the global, zero-form symmetry defined on the vertex sub-lattice $A$, which is generated by $\prod_{v\in A} \tau_v^x$; it can be viewed as an extension of the $\mathbb{Z}_2$ symmetry in one-dimensional cluster states. Meanwhile, $\mathbb{Z}_2^{(1)}$ is a subsystem, one-form symmetry defined for a loop on the edge sub-lattice $B$  \cite{yoshida2016generalized,jian2021higher,you2018subsystem}, which is generated by $\prod_{e\in\text{loop on }B}\sigma_e^x$, leading to an exponentially large number of the conserved charges in the thermodynamic limit (see Fig.~\ref{fig:main}(a)).

A smoking gun of a pure SPT state is the emergence of boundary degrees of freedom associated with the ground-state degeneracy under open boundary conditions. We argue that this defining feature should continue to be crucial in diagnosing mixed-state SPT phases, even though the concept of the ground state becomes ambiguious. In the case of a pure 2D cluster state with a boundary  $\partial \mathcal{M}$, there exist $2^{\lvert \partial \mathcal{M} \rvert}$ (nearly) degenerate ground states, where $\lvert \cdot\rvert$ represents the perimeter. 
To define the corresponding boundary spin-$\frac{1}{2}$ operators  that commute with $H_{\text{Cluster}}$, we can introduce the following operators on the boundary vertices $u$ (see Fig.~\ref{fig:main}(b)): 
\begin{align}
&\pi^x_{u}=\tau^x_{u}\sigma^z_{e_1}\sigma^z_{e_2}\sigma^z_{e_3},\qquad\pi^y_{u}=\tau^y_{u}\sigma^z_{e_1}\sigma^z_{e_2}\sigma^z_{e_3},\nonumber\\
&\pi^z_{u}=\tau^z_{u},
\end{align}
where $\boldsymbol{\pi}_{u}=(\pi^x_{u},\pi^y_{u},\pi^z_{u})^{\rm T}$ obeys the algebraic relations of the spin-$\frac{1}{2}$.

To proceed with our discussion, we need to specify a particular ground state. To be concrete, we choose a ground state where $\pi^x_{u}=1$ for all  $u$,
\begin{gather}
\rho_{\text{SPT}}=\prod_{u\in \partial \mathcal{M}}\frac{1+\pi^x_{u}}{2}\prod_{v\in A}\frac{1+A_v}{2}\prod_{e\in B}\frac{1+B_e}{2},
\end{gather}
while we emphasize that its choice does not affect our main results; another ground state can be obtained simply by performing a boundary flip with $\pi^z_{u}=\tau^z_{u}$. Since we have the relation $B_e=1$ within the ground-state subspace, we can define the following string operator $\mathcal{S}_{u,u'}$ that simultaneously flips two boundary spins on $u$ and $u'$:
\begin{gather}
\label{String}
\mathcal{S}_{u,u'}=\prod_{e\in {\gamma_{uu'}}} \sigma^x_e,
\end{gather}
where ${\gamma}_{uu'}$ represents a string on sub-lattice $B$ that connects through $\sigma$ operators between the boundary spins $\pi_{u}$ and $\pi_{u'}$. Importantly, this operator satisfies 
\eqn{
\text{tr}[\rho_{\text{SPT}}\mathcal{S}_{u,u'}\rho_{\text{SPT}}\mathcal{S}_{u,u'}]=0,
}
which means that energetically degenerate SPT states with different edge states are orthogonal to each other. 

\section{Effects of local decoherence}
In this section, we model decoherence as the combination of two local quantum channels which describe the single-qubit flip or phase error:
\eqn{
\mathcal{N}^\alpha_i[\rho]=\begin{cases}
 (1-p_\alpha)\rho+p_\alpha\tau_i^\alpha\rho\tau_i^\alpha & i\in A\\
 (1-p_\alpha)\rho+p_\alpha\sigma_i^\alpha\rho\sigma_i^\alpha & i\in B\\
\end{cases}
}
with $\alpha\in\{x,z\}$. Here, $p_x$ and $p_z$ characterize the spin-flip and phase decoherence rates, respectively. For the sake of simplicity, we assume that these rates take the same values on both sub-lattices $A$ and $B$. The resulting mixed  state reads as:
\begin{gather}
\rho_D=\prod_i \mathcal{N}^Z_i\circ\mathcal{N}^X_i[\rho_{\text{SPT}}].
\end{gather}

\begin{figure}[t]
    \centering
    \begin{minipage}[t]{0.22\textwidth}
        \centering
      \includegraphics[width=\textwidth]{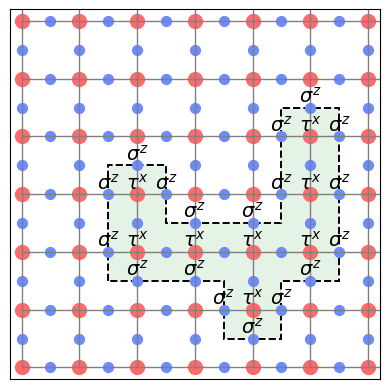}
    \end{minipage}
    \hfill
    \begin{minipage}[t]{0.22\textwidth}
        \centering
   \includegraphics[width=\textwidth]{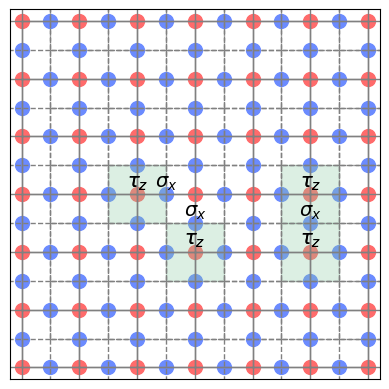}
    \end{minipage}
    \caption{(Left) Schematic of the domain-wall (Dw) configuration $\rho_{\text{Dw}}$ where Pauli operators are explicitly shown while identities are represented by unlabeled markers. The green zone labels a domain wall. (Right)  Schematic of gauge (Ga) configuration $\rho_{\text{Ga}}$. The green zone represents the $\mathbb{Z}_2$ flux with $ \phi_\square=-1 $.}
    \label{fig:2}
\end{figure}

In the rest of this section, we will present $\rho_D$ within the framework of the domain-wall condensing picture by utilizing a summation over domain configurations. To this end, we represent each of the products resulting from the expansion of $\prod_{v\in A}(1+A_v)$ by a certain domain wall (Dw). Specifically, suppose that either $\tau^x$ or identity operator is assigned to every vertex on $A$. If an edge connects two vertices, one carrying a $\tau_x$ and the other an identity operator, we place $\sigma_z$ on this edge. Each corresponding configuration is then associated with $\rho_{\text{Dw}}$ (see Fig.~\ref{fig:2}(a)). As such, we can interpret this problem in terms of a classical spin variable $s$ by identifying $\tau^x$ as $s=+1$ and the identity operator as $s=-1$. 

Similarly,  each of the products obtained by expanding $\prod_{e\in B} (1+B_e)$ can be considered as a certain gauge configuration (Ga) on the dual lattice. Namely, either $\sigma^x$ or identity operator is assigned to every edge on $B$, and if there are an odd number of edge qubits surrounding a vertex qubit, we place a $\tau^z$ on the vertex. Identifying each $\sigma^x$ (identity operator) on $B$ as $s=1$ ($s=-1$), we can interpret the corresponding contribution $\rho_{\text{Ga}}$ in terms of the gauge configuration where the product of four classical spins $\{s_i\}$ on the square $\square$ surrounding a vertex qubit $\tau_z$ is constrained to be $\phi_{\square}=\prod_{\square} s_i=-1$ (see Fig.~\ref{fig:2}(b)).

The density matrix for the pure state then reads as:
\begin{gather}
    \rho_{\text{SPT}}=\frac{1}{2^{3N^2}}\sum_{\text{Dw}}\rho_{\text{Dw}}
    \sum_{\text{Ga}}\rho_{\text{Ga}},
\end{gather}
where the summations run over all possible domain-wall and gauge configurations, and every term has the same coefficient. 
The local quantum channels only change the coefficients before $\rho_{\text{Dw}}$ and $\rho_{\text{Ga}}$. Specifically, the bit-flip error is given by
\eqn{
    \mathcal{N}^X_i[\rho_{\text{Dw}}]&=&\begin{cases}
        (1-2p_x)\rho_{\text{Dw}}& \text{{$\sigma^z_i\in$ domain wall}}\\
        \rho_{\text{Dw}}  & {\text{otherwise}    }
    \end{cases}\\\nonumber
         \mathcal{N}^X_i[\rho_{\text{Ga}}]&=&\begin{cases}
        (1-2p_x)\rho_{\text{Ga}}&\text{$\tau^z_i\in$ vertex with flux $-1$}\\
        \rho_{\text{Ga}}   & {\text{otherwise}    }   
    \end{cases} \nonumber\\
}
while the phase error is
\eqn{
    \mathcal{N}^Z_i[\rho_{\text{Dw}}]&=&\begin{cases}
        (1-2p_z)\rho_{\text{Dw}}&\text{{$s_i=1$ on vertex}}\\
        \rho_{\text{Dw}}  & {\text{otherwise}    }    
    \end{cases}\\
        \mathcal{N}^Z_i[\rho_{\text{Ga}}]&=&\begin{cases}
        (1-2p_z)\rho_{\text{Ga}}&\text{{$s_i=1$ on edge}}\\
        \rho_{\text{Ga}}      & {\text{otherwise}    }
    \end{cases}.
}
As a result, the decohered mixed state can be represented as a superposition of operators where the coefficient before each operator is determined from the corresponding classical spin configurations. To see this, we can factorize $\rho_D$ into the contributions associated with each of sub-lattices denoted by $\rho_D^A$ and $\rho_D^B$ as follows:
\eqn{
\rho_D^A&=&\frac{1}{2^{N^2}}\sum_{\text{Dw}}(1-2p_x)^{\lvert\partial\mathcal{D}\rvert}(1-2p_z)^{\lvert s=+1\rvert}\rho_{\text{Dw}},\nonumber\\
\rho_D^B&=&\frac{1}{2^{2N^2}}\sum_{\text{Ga}}(1-2p_x)^{\lvert \phi=-1\rvert}(1-2p_z)^{\lvert s=+1\rvert}\rho_{\text{Ga}}.\nonumber
}
Here, $\lvert\partial\mathcal{D}\rvert$ represents the length of the domain wall, $\lvert s=+1\rvert$ is the number of classical spins having $s=+1$, and $\lvert \phi=-1\rvert$ is the number of fluxes having $\phi=-1$. The coefficients for each configuration can be related to statistical weights of a classical spin model. Specifically, we have
\eqn{
    \rho_D&=&\rho_D^A\rho_D^B\nonumber\\
    &\propto&\sum_{\text{Dw}} e^{-H_{\text{Ising}}/2} \rho_{\text{Dw}} \sum_{\text{Ga}} e^{-H_{\text{Gauge}}/2} \rho_{\text{Ga}},
    \label{eq11}
}
where both summations run over all the possible classical spin configurations on a square lattice, and $H_{\text{Ising}}$ and $H_{\text{Gauge}}$ are given by
\eqn{\label{ising}
H_{\text{Ising}}&=&-J\sum_{\langle ij \rangle}s_is_j-h\sum_{i}s_i,\\
H_{\text{Gauge}}&=&-U\sum_{\square}s_is_js_ks_l-t\sum_{i}s_i.\label{isgauge}
}
Here, we define the coupling strengths by $J=-\ln{(1-2p_x)}$, $h=-\ln{(1-2p_z)}$, $U=-\ln{(1-2p_x)}$, and $t=-\ln{(1-2p_z)}$. Consequently, the second-order R\'enyi entropy of the decohered mixed state can be obtained by
\begin{gather}
\tr \rho_{D}^2=\tr (\rho^{A}_D)^2\;\tr (\rho^{B}_D)^2\propto\mathcal{Z}_\text{Ising}\mathcal{Z}_\text{Gauge},
\end{gather}
where $\mathcal{Z}_\text{Ising}$ and $\mathcal{Z}_\text{Gauge}$ are the partition functions of $H_{\text{Ising}}$ and $H_{\text{Gauge}}$, respectively. It is well-known that the 2D Ising gauge theory~\eqref{isgauge} does not exhibit a phase transition because its Wilson loop operator always displays an area law behavior, $e^{-W[C]}\sim e^{-\text{Area of }C}$. As such, nonanalytic behavior of the second-order Rényi entropy, if any, should be attributed to the partition function of the Ising model, which undergoes a phase transition between PM and FM phases at $p_x=p^{(2)}_c=\frac{1-\sqrt{\sqrt{2}-1}}{2}\sim0.1782$ 
and $p_z=0$. 
{Said differently, this observation indicates that even arbitrarily weak phase error $p_z>0$ can break the SPT order. In fact, our analysis is consistent with the fact that no SPT order is expected when a state is protected solely by the average symmetries, which are satisfied only after taking the ensemble average over quantum trajectories \cite{ChongWang2023Average} (see the discussion below and Appendix~\ref{Ap1} for further details).}
Since our primary focus is on the decoherence-induced topological phase transition, we shall assume $p_z=0$ and focus on the effects of the bit-flip error from now on. 

It is noteworthy that higher-order R\'enyi entropies could also be factorized and mapped to classical partition functions as follows:
\begin{gather}
    \tr\rho_D^n=\tr(\rho_D^{A})^n\;\tr(\rho_D^{B})^n\propto\mathcal{Z}^n_A\mathcal{Z}^n_B,
\end{gather}
where $\mathcal{Z}^n_A$ is the partition function of the $(n-1)$-flavor Ising model
\begin{gather}\label{multiising}
    H_n=-\frac{J_n}{2}\sum_{\langle ij \rangle}(\sum_{\alpha=1}^{n-1}s_i^{(\alpha)}s_j^{(\alpha)}+\prod_{\alpha=1}^{n-1}s_i^{(\alpha)}s_j^{(\alpha)}),
\end{gather}
and
$\mathcal{Z}^n_B$ is the partition function of the $(n-1)$-flavor Ising gauge model, 
\eqn{H'_n=-\frac{U_n}{2}\sum_{\square}(\sum_{\alpha=1}^{n-1}\phi_{\square}^{(\alpha)}+\prod_{\alpha=1}^{n-1}\phi_{\square}^{(\alpha)}),} 
which exhibits no phase transition \cite{Kogut1979}. 
The critical decoherence error rate $p_c^{(n)}$ associated with the $(n-1)$-flavor Ising model~\eqref{multiising} monotonically increases with the replica index $n$ \cite{KM81,fan2023diagnostics,bao2023mixedstate}. In particular, in the limit $n\to\infty$, the model becomes a solvable decoupled Ising model, for which $p^{(\infty)}_c=\frac{2-\sqrt{2}}{2}\sim0.2929$. A quantum-information interpretation of this increasing sequence of critical error rates is that, while qubit errors in general degrade the computational power of the cluster state, more copies of the system can still provide available resources. Nevertheless, there is ultimately a fundamental limit on the error rate, $p^{(\infty)}_c$, above which no matter how many copies of the mixed state are available, one cannot extract any quantum computational resource from them.

We here comment on the symmetries of the decohered mixed state. Namely, in order to define a SPT phase in open quantum systems, one must preserve all the relevant symmetries in a certain sense even in the context of a mixed state. There are two ways to define a unitary symmetry for a density matrix. The first way is the so-called  exact or strong symmetry condition denoted by $\rho_D=U(g_L)\rho_DU^\dagger(g_R)$, where $U(g_{L,R})$ are the unitary operators associated with $g_{L,R}\in G$. This is the direct generalization of the symmetry condition for a quantum state and guarantees that the symmetry is satisfied for every individual quantum trajectory. The second one is referred to as the  average or weak symmetry condition, which is represented as $\rho_D=U(g)\rho_DU^\dagger(g)$ with $g\in G$ \cite{ChongWang2023Average, ma2023topological}, and this is the condition satisfied after taking the ensemble average over all trajectories. One can check that the mixed state of our primary focus, namely, $\rho_D$ with $p_z=0$ and $p_x>0$, satisfies the exact $\mathbb{Z}_2^{(0)}\times\mathbb{Z}_{2}^{(1)}$ symmetry, while nonzero $p_z$ immediately renders both symmetries average ones.  We note that a decohered SPT state with an exact-average mixed symmetry, such as $\mathbb{Z}_{2,\text{avg}}^{(0)}\times\mathbb{Z}_{2}^{(1)}$ or $\mathbb{Z}_{2}^{(0)}\times\mathbb{Z}_{2,\text{avg}}^{(1)}$, have been studied in terms of strange correlation functions \cite{lee2022symmetry} and separability \cite{chen2023symmetry}. We refer to Appendix~\ref{Ap1} for further details about how different quantum channels correspond to distinct symmetry conditions.

\label{Sec3}
\section{Diagnostics of topological phase transition}
In the previous section, we discuss a phase transition in the R\'enyi entropy of the decohered mixed state. However, it remains unclear whether or not this nonanalytic behavior is accompanied by a topological phase transition. In this section, we present three diagnostic methods for identifying the nature of the SPT phase transition. In particular, we show that the PM to  FM phase transition found in the second-order R\'enyi entropy indeed corresponds to the topological phase transition, where the PM (FM) phase in the classical spin model is the SPT (trivial) phase in the quantum problem. Furthermore, diagnostic tests based on higher-order R\'enyi entropies and replica limit will also be discussed.
\label{Sec4}

\subsection{R\'enyi relative entropy}
The existence of nontrivial boundary states is one of the defining features of SPT phases. As a diagnostic test for SPT states in open quantum systems, we here examine the fate of the boundary degrees of freedom in the presence of decoherence. This can be achieved by measuring the difference between the decohered mixed state $\rho_D=\mathcal{N}[\rho_{\text{SPT}}]$ and $\rho_{S}=\mathcal{N}[\mathcal{S}_{u,u'}\rho_{\text{SPT}}\mathcal{S}_{u,u'}^{-1}]$, where we recall that $\mathcal{S}_{u,u'}$ is the string operator that flips two boundary spins $\pi$ at $u$ and $u'$; see Eq.~(\ref{String}). For this purpose, we introduce the R\'enyi relative entropy between $\rho_D$ and $\rho_S$:
\begin{gather}
D^n(\rho_D\lvert\lvert\rho_S)=\frac{1}{1-n}\ln\frac{\tr\rho_D\rho_S^{n-1}}{\tr \rho^{n}_D}.\label{renyin}
\end{gather}
When $n$ is taken to be 1, Eq.~\eqref{renyin} reproduces the von Neumann relative entropy, $D^1(\rho_D\lvert\lvert\rho_S)=S(\rho_D\lvert\lvert\rho_S)=\tr\rho_D(\ln{\rho_D}-\ln{\rho_S})$. 
In the absence of decoherence, $D^n(\rho_D\lvert\lvert\rho_S)$ diverges since $\rho_D$ and $\rho_S$ are orthogonal pure states. Under weak decoherence, we expect the boundary spins to be less localized {compared to those of the pure SPT state}, and flipping two boundary spins nearby would have negligible impact on the system. In contrast, when $\lvert i-j\rvert\gg 1$, the boundary degrees of freedom should still be distinguishable, and $D^n(\rho_D\lvert\lvert\rho_S)$ is expected to diverge as $\lvert i-j\rvert$ goes to infinity. Meanwhile, in the topologically trivial phase without nontrivial boundary spins, no matter how large  $\lvert i-j\rvert$ is, the relative entropy should give a finite value. These distinct long-distance behaviors of the relative entropy can be used as a diagnosis of the topological phase transition. We note that such disappearance of distinguishable boundary states can be interpreted as a computational power phase transition in the context of MBQC since the target qubits are decoding on the boundary of the 2D cluster state. 

Intriguingly, the R\'enyi relative entropy~\eqref{renyin} can be mapped to the logarithm of the boundary spin-spin correlation function in the 2D Ising model. To prove the mapping, we firstly write down the domain-wall expansion of $\rho_S$:
\begin{gather}
    \rho_S=\frac{1}{2^{N^2}}\sum_{\text{Dw}} (-1)^{\#\mathcal{S}\cap\text{Dw}} e^{-H_{\text{Ising}}/2}\rho_{\text{Dw}}\rho^B_D,
\end{gather}
where $\#\mathcal{S}\cap\text{Dw}$ counts how many times the string operator $\mathcal{S}_{u,u'}$ crosses the domain wall, which accounts for the sign difference between the boundary spins $s_u$ and $s_{u'}$. 
We then get
\begin{gather}
\tr\rho_D\rho_S^{n-1}\propto\mathcal{Z}^n_B\prod_{\alpha=2}^n\sum_{\text{Dw}^{(\alpha)}}(-1)^{\#\mathcal{S}\cap\text{Dw}^{(\alpha)}}e^{-H_n},
\end{gather}
where $\alpha$ is the replica index, and we use the fact that only the terms satisfying $\prod_{\alpha=1}^n\rho_{\text{Dw}}^{(\alpha)}=1$ contribute to the trace. Using the relation $\prod_{\alpha=2}^n(-1)^{\#\mathcal{S}\cap\text{Dw}^{(\alpha)}}=(-1)^{\#\mathcal{S}\cap\text{Dw}^{(1)}}=s^{(1)}_u s^{(1)}_{u'}$, aside an irrelevant constant term we obtain
\eqn{
    D^n(\rho_D\lvert\lvert\rho_S)&=&\frac{1}{1-n}\ln \frac{1}{\mathcal{Z}^n_A}\sum s^{(1)}_is^{(1)}_je^{-H_n}\nonumber\\
    &=& \frac{1}{1-n}\ln \langle s^{(1)}_i s^{(1)}_j\rangle_{\text{Boundary}}.
}
Here, we note that the string operator commutes with any term in the gauge configuration, and the Ising gauge theory does not contribute to the final result. Since the spin-spin correlation function decays exponentially in the  PM phase, the relative entropy $D^n(\rho_D\lvert\lvert\rho_S)$ is proportional to the distance $\lvert i-j\rvert$ between $i$ and $j$, indicating that the mixed state retains distinguishable boundary states. In the FM phase, in contrast, the correlation function acquires a distance-independent constant due to the long-range order, and the relative entropy has a finite constant even if $\lvert i-j\rvert\gg 1$, indicating that the mixed state loses its boundary degrees of freedom.

\subsection{Strange correlation function}
Detecting SPT phases without open boundaries is rather nontrivial since SPT phases are symmetry-preserved gapped systems with only short-range entanglement. One possible way for this is to use the strange correlation function. Specifically, for a $G$-SPT phase, the strange correlation function is defined by
\begin{gather}
C=\frac{\text{tr}\rho_0\mathcal{O}\rho}{\text{tr}\rho_0\rho},
\end{gather}
where $\rho$ is a state to be detected, and $\rho_0=\ket{\Omega}$ with $\bra{\Omega}$ being a product state. The operator $\mathcal{O}$ itself carries a nontrivial $G$ charge or is the correlation of operators carrying nontrivial $G$ charge. The state $\rho$ is trivial if the correlation function decays quickly, while $\rho$ is topological if the correlation function saturates to a constant.

In our model, the symmetry $G$ is $\mathbb{Z}_2^{(0)}\times\mathbb{Z}_2^{(1)}$; for the sake of concreteness, we shall focus on the decoherence effect on the one-form symmetry $\mathbb{Z}_2^{(1)}$ here. The one-form charge or one-form defect is carried by a string operator defined as $\mathcal{O}=\prod_{e\in\gamma}\sigma^z_e$, where $\gamma$ is a loop on sub-lattice $B$. The $\mathbb{Z}_2^{(1)}$ charge associated with $\cal O$ is trivial (nontrivial) if the string $\gamma$ intersects with loop-like symmetry operators an even (odd) number of times. As the reference product state, we choose $\ket{\Omega}=\ket{1}^{\otimes 2N^2}\ket{+}^{\otimes N^2}$, where all the qubits on sub-lattice $A$ ($B$) satisfy $\tau_x\ket{+}=\ket{+}$ ($\tau_z\ket{1}=\ket{1}$). The strange correlation function of our model is then given by
\begin{gather}\label{stcorr}
    C(\gamma)=\frac{\text{tr}\rho_0\prod_{e\in\gamma}\sigma^z_e\rho_D}{\text{tr}\rho_0\rho_D},
\end{gather}
which has an alternative expression in the corresponding classical model by
\begin{gather}
\label{eq23}
 C(\gamma)=e^{\mathcal{F}_{\text{Ising}}-\mathcal{F}_{\text{Ising}/\gamma} }.
\end{gather}
Here, $\mathcal{F}_{\text{Ising}/\gamma}$ is the free energy of the modified Ising model in which all the bonds crossing loop $\gamma$ are cut. This result can again be proved by the domain-wall expansion. To do so, we firstly move the quantum channel from $\rho_{\text{SPT}}^{\text{(PBC)}}$ to $\rho_0$ as follows:
\eqn{
    \tr\rho_0\rho_D&=&\tr \mathcal{N}[\rho_0]\rho_{\text{SPT}}^{\text{(PBC)}}\nonumber\\
    &=&\bra{\Psi_{\text{SPT}}}\prod_{e\in B}\frac{1+(1-2p_x)\sigma^z_e}{2}
   \nonumber\\
   & & \times\prod_{v\in A}\frac{1+\tau^x_v}{2}\ket{\Psi_{\text{SPT}}},
}
where $\ket{\Psi_{\text{SPT}}}$ is the unique ground state of the stabilizer Hamiltonian under periodic boundary conditions. Since only the configurations where $\sigma^z$'s form a closed loop and its inside is filled with $\tau^x$'s contribute to the final result, we obtain
\begin{gather}
    \tr\rho_0\rho_D=\frac{1}{2^{3N^2}}\sum_{\text{Dw}}(1-2p_x)^{{|\partial{\cal D}|}}\propto\mathcal{Z}_{\text{Ising}}.
\end{gather}
Similarly, we can rewrite the numerator of Eq.~\eqref{stcorr} as
\begin{gather}
    \tr\rho_o\mathcal{O}\mathcal{N}[\rho_{\text{SPT}}]=\nonumber\\ \bra{\Psi_{\text{SPT}}}\prod_{e\in\gamma}\frac{1+\sigma^z_e}{2}\prod_{e\in B/\gamma}\frac{1+(1-2p)\sigma^z_e}{2}\prod_{v\in A}\frac{1+\tau^x_v}{2}\ket{\Psi_{\text{SPT}}},
\end{gather}
where $B/\gamma$ denotes sub-lattice $B$ without loop $\gamma$. If the domain wall does not intersect with $\gamma$, the probability of the domain wall appearing is related to the perimeter of the domain wall, which corresponds to a coupling strength $-\ln (1-2p_x)$ in the Ising model. However, if the domain wall intersects with the closed loop $\gamma$, the part overlapping with $\gamma$ acquires a factor of $(1-2p_x)$. This effectively sets the Ising coupling strength to be 0; in other words, it divides the Ising model along $\gamma$ into two parts. Hence, we have 
\eqn{
\tr\rho_0\mathcal{O}\mathcal{N}[\rho_{\text{SPT}}]\propto\mathcal{Z}_{\text{Ising}/\gamma},} 
which proves Eq.~(\ref{eq23}).
\begin{figure}[b]
    \centering
   \includegraphics[width=0.45\textwidth]{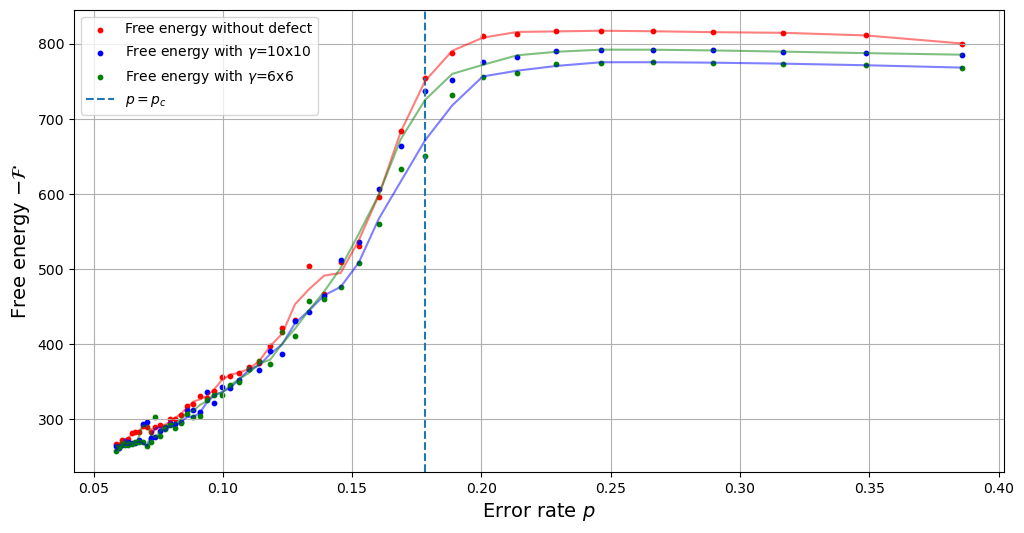}
    \caption{Monte Carlo computation of free energy is illustrated both with and without bond crossing \( \gamma \). The red curve represents the free energy of the original Ising model. The green and blue curves represent the Ising model with loop defects of sizes \( \gamma = 10 \times 10 \) and \( \gamma = 6 \times 6 \) respectively. Simulations were conducted on a \( 20 \times 20 \) lattice with $10^5$ sampling iterations. There exist noticeable free-energy differences in the FM phase with \( p_x \) being high, while the free-energy excesses vanish in the PM phase with a small error rate.  }
    \label{monte}
\end{figure}

The consequence of Eq.~(\ref{eq23}) is readily apparent; in the deep PM phase, the leading order of the free energy change is $\Delta\mathcal{F}=o(p^2_x)$, and $C(\gamma)$ is almost insensitive to the length of the loop, which is denoted by $\lvert\gamma\rvert$. In the FM phase, in contrast, the Ising interaction contributes to the free energy of this model,  and the leading order becomes $\Delta\mathcal{F}=O(\lvert\gamma\rvert)$. Consequently, $C(\gamma)$ decays exponentially as $|\gamma|$ is increased. We also demonstrate this by performing a classical Monte Carlo stimulation of the free-energy excess in the Ising model as shown in Fig.~\ref{monte}.

\subsection{Multipartie negativity}
\begin{figure}[b]
   \includegraphics[width=0.30\textwidth]{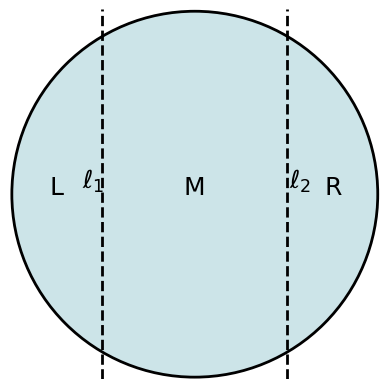}
    \caption{Cuttings of a 2D disk into three parts \( L,M \) and \( R \) where \( L \) and \( R \) are well separated. \( l_1 \) and \( l_2 \) label the lengths of cutting lines. }
    \label{fig:cut}
\end{figure}

As an alternative diagnosis of SPT phases in open quantum systems, we finally discuss the multipartite entanglement entropy, which is defined by \( I(L;R|M)=S_{LM}+S_{MR}-S_{M}-S_{LMR} \) where regions \( L \) and \( M \) are sufficiently separated; see Fig.~\ref{fig:cut}. References \cite{zeng2015gapped,zeng2016topological,fromholz2020entanglement} have proposed that nonzero values of \( I(L;R|M) \) can serve as a means to detect nontrivial many-body entanglement and distinguish between SPT and trivial phases. It would be interesting to calculate an entanglement measure similar to the multipartite entanglement entropy in the present setup if at all possible. For this purpose, we introduce a multipartite negativity denoted by \( N(L;R|M) \), which should serve as an open-system analog of \( I(L;R|M) \). Below we demonstrate that the tripartite negativity \( N(L;R|M) \) exhibits the singular behavior similar to that of \( I(L;R|M) \) across the topological phase transition. We note that Refs.~\cite{wen2016edge,wen2016topological,shapourian2017partial,lu2020detecting,lu2022characterizing,lavasani2021measurement} have also utilized the negativity to study the topological phases of matter.

The negativity of a subsystem \( L \) is defined as follows \cite{vidal2002computable,horodecki1996necessary,Plenio05Negativity}:
\begin{gather}
\mathcal{E}_L(\rho)=\ln\lVert \rho^{T_L} \rVert_1,
\end{gather}
where \( T_L \) represents the partial transpose of all degrees of freedom in subsystem \( L \), and \( \lVert \cdot \rVert_1 \) is the trace norm. Additionally, we consider a series of R\'enyi negativities
\begin{gather}
\mathcal{E}_L^{(2n)}(\rho)=\ln \frac{\text{Tr} (\rho^{T_L})^{2n} }{\text{Tr} \rho^{2n}},
\end{gather}
which reduce to the trace-norm negativity \( \mathcal{E}_L \) as \( n \) approaches 1 \cite{calarbrese2012negativity}.
The R\'enyi tripartite negativity is then defined by
\eqn{
N^{(2n)}(L;R|M)=\mathcal{E}_{LM}^{(2n)}+\mathcal{E}_{MR}^{(2n)}-\mathcal{E}_{M}^{(2n)}-\mathcal{E}_{LMR}^{(2n)},}
which serves as our detector for nontrivial topological phases. We here note that, in general, phases with nonzero \( I(L;R|M) \) or \( N(L;R|M) \) can belong to either topological or symmetry-breaking phases \cite{zeng2019quantumbook}. Nevertheless, in our mixed state the symmetry is always preserved, and nonzero information quantities must always indicate the topological phases.

In a decohered mixed state, \( \mathcal{E}_L^{(2n)} \) can be given by the free-energy excess in the corresponding classical model where a constraint is added on the Ising model such that no domain walls are allowed to cross the boundary of subregion \( L \): 
\begin{gather}\label{multin}
    \mathcal{E}_L^{(2n)}=\Delta\mathcal{F}_L.
\end{gather}
To prove the relation, we again expand the density matrix in the domain-wall picture:
\begin{gather}
    \rho_D^{T_L}=\sum_{\text{Dw}}\sum_{\text{Ga}}(-1)^{y(\rho_{\text{Dw}}\rho_{\text{Ga}})}\rho_{\text{Dw}}\rho_{\text{Ga}}e^{-H_{\text{Gauge}}/2-H_{\text{Ising}}/2},
\end{gather}
where \( {y(\rho_{\text{Dw}}\rho_{\text{Ga}})} \) is the number of the Pauli \( Y \) operators in the region \( L \), and we use the fact that only the Pauli \( Y \) operators contribute to the partial transpose. For the sake of simplicity, in the rest of this section, we set the error rates of sub-lattice \( L \) to zero or, equivalently, the interaction strength \( U \) in \( H_{\text{gauge}} \) to zero, which allows us to calculate the summation exactly; we note that this assumption does not affect our conclusions that are independent of the details of the Ising gauge partition function. Accordingly, we get 
\begin{gather}
    \tr(\rho_D^{T_L})^n\!=\!\sum_{\text{Dw}}\sum_{\text{Ga}}(-1)^{\prod_{\alpha=1}^{n-1}y(\rho^{(\alpha)}_{\text{Dw}}\rho^{(\alpha)}_{\text{Ga}})y(\prod_{\alpha=1}^{n-1}\rho^{(\alpha)}_{\text{Dw}}\rho^{(\alpha)}_{\text{Ga}})} e^{-H_n},
\end{gather}
To rewrite the expression, we introduce \( {\rm sgn}(\rho_1,\rho_2)=\pm 1 \) depending on the commutation relation of \( \rho_1 \) and \( \rho_2 \) in the subregion \( L \), where \( {\rm sgn}(\rho_1,\rho_2)=1 \) if \( \rho_1 \) commutes with \( \rho_2 \); otherwise \( {\rm sgn}(\rho_1,\rho_2)=-1 \). Utilizing the fact that \( (-1)^{y(\rho_1\rho_2)}=(-1)^{y(\rho_1)}(-1)^{y(\rho_2)}{\rm sgn}(\rho_1,\rho_2) \) and all \( \rho_{\text{Dw}} \) or \( \rho_{\text{Ga}} \) commute with each other, we have 
\begin{gather}
    \tr(\rho_D^{T_L})^n=\sum_{\text{Dw}}\sum_{\text{Ga}}{\prod_{\alpha=1}^{n-1}\prod_{\alpha'=1,\alpha'\neq \alpha}^{n-1}}{\rm sgn}(\rho_{\text{Dw}}^{(\alpha)},\rho_{\text{Ga}}^{(\alpha')})e^{-H^n}.
\end{gather}
The summation \( \sum_{\text{Ga}}{\prod_{\alpha=1}^{n-1}\prod_{\alpha'=1,\alpha'\neq \alpha}^{n-1}}{\rm sgn}(\rho_{\text{Dw}}^{(\alpha)},\rho_{\text{Ga}}^{(\alpha')}) \) can be done by noticing that \( \rho_{\text{Dw}} \) and \( \rho_{\text{Ga}} \) always commute. So \( \rm sgn \) will always be \( +1 \) if \( \rho_{\text{Ga}} \) does not cross the boundary of \( L \); otherwise the sign fluctuation will force the summation to be zero. As a result, we get 
\begin{gather}
\sum_{\text{Ga}}{\prod_{\alpha=1}^{n-1}\prod_{\alpha'=1,\alpha'\neq \alpha}^{n-1}}{\rm sgn}(\rho_{\text{Dw}}^{(\alpha)},\rho_{\text{Ga}}^{(\alpha')})\propto\prod_{\alpha=1}^{n-1} \delta(\rho_{\text{Dw}}^{(\alpha)},L).
\end{gather}
We note that \( \delta(\rho_{\text{Dw}}^{(s)},L) \) is equal to 1 if no domain walls cross the boundary of region \( L \). This can be interpreted as a constraint on the Ising model. Denoting the partition function of the Ising model with this constraint by \( \mathcal{Z}'_{n} \), we finally obtain \( \mathcal{E}_L^{(2n)}(\rho)=\ln \frac{\mathcal{Z}'_{2n}}{\mathcal{Z}_{2n}}\equiv\Delta{\cal F}_L \), which proves Eq.~\eqref{multin}.

In our model, the negativity of the entire system consistently remains zero, i.e., \( \mathcal{E}_{LMR}^{(2n)}=0 \). Additionally, Refs.~\cite{metlitski2009entanglement,fan2023diagnostics} have argued that the leading-order in the excess free energy \( \Delta {\cal F}_L \) is directly proportional to the length of \( \partial L \) in both PM and FM phases for a large enough subregion \( L \) satisfying \( \lvert\partial L\rvert\gg \xi \), where \( \xi \) is the correlation length \footnote{It is important to note here that \( |\partial L| \) should not include the length of the boundary of the entire system.}. The leading contribution thus satisfies the area law; for instance, \( \mathcal{E}_{L}^{(2n)}=cl_1 \) and \( \mathcal{E}_{M}^{(2n)}=c(l_1+l_2) \), where \( l_1 \) (\( l_2 \)) is the perimeter of the boundary between \( L \) and \( M \) (\( M \) and \( R \)) and \( c \) is some constant \footnote{Here, $l_1$ and $l_2$ should have the same coefficient $c$ when $l_i$ $\gg$ $\xi$ and their distance is also much greater than $\xi$, in which the coefficient is not expected to depend on the geometry}. 
Meanwhile, the topological nature manifests itself as the nontrivial subleading contribution to the excess free energy. Namely, in the PM phase, there is an additional crucial contribution from the boundary entropy, which is \( -\ln 2 \), as the spins on the boundary can be either up or down. In the FM phase, however, all spins must align with the same direction since it is not allowed to have a fluctuating domain wall when the size of domain wall is larger than the correlation length \( \xi \).  All in all, we finally arrive at the following conclusion:
\begin{gather}
    N^{(2n)}(L;R|M)=\begin{cases} 0 &\text{FM phase}\\
    \ln2&\text{ PM phase}   
    \end{cases},
\end{gather}
{which implies that the multipartite negativity can indeed play the role as a probe of the SPT order in mixed states, in a similar manner as the multipartite entanglement entropy does in the case of pure states.}

\section{conclusion\label{Sec5}}
In this article, we explore a topological phase transition of a 2D cluster influenced by local decoherence. We find that the 2D cluster state under the bit-flip error can represent a SPT phase protected by the exact or strong $\mathbb{Z}_2^{(0)}\times\mathbb{Z}_{2}^{(1)}$ symmetry. This work, combined with previous studies\cite{chen2023symmetry,lee2022symmetry}, contributes to a systematic understanding of how 2D cluster states endure under decoherence and aid the transition between pure-state SPT and average SPT orders. Furthermore, we map the decohered cluster state to Ising-like models, which significantly simplifies the calculations of various SPT diagnostics motivated by different perspectives, translating the original quantum problems into analytically tractable statistical mechanical models.

{Our model demonstrates the transition of 2D bosonic SPT order under the influence of local quantum errors. The theory underlying such states is of particular importance, as it attracts significant interest in the context of noisy intermediate-scale quantum platforms, which aim to realize various topological phases even in the absence of quantum error correction. There, it will be crucial to identify operational quantities that are nonlinear functions of the density matrix. It is also natural to explore whether or not there exist critical behaviors beyond the currently known universal classes. Moreover, the development of a quantized entanglement index \cite{Kim2022Chiral,Gong2021Topological,Fan2023Chern} for detecting SPT order in open systems remains an open area of research.

Another exciting avenue for further exploration is the realm of decohered fermionic systems. It merits further study, for instance, to explore the behavior of free fermions in the 10-fold way \cite{Ryu:2010fe} under decoherence. It is plausible to hypothesize the emergence of diverse symmetry-preserved decoherence channels for topological insulators and superconductors. Meanwhile, the original gapless topological phase transition points may also change by decoherence in a nontrivial manner, potentially leading to a richer understanding of critical behaviors in open quantum systems  
 \cite{Murciano2023Measurement,PRXQuantum.4.030317}.

\begin{acknowledgments}
We are grateful to Hosho Katsura for useful discussions. We thank Ryotaro Niwa for collaboration on a related project. Y.A. acknowledges support from the Japan Society for the Promotion of Science (JSPS) through Grant No. JP19K23424 and from JST FOREST Program (Grant Number JPMJFR222U, Japan) and JST CREST (Grant Number JPMJCR23I2, Japan).
\end{acknowledgments}

\appendix
\section{Symmetry conditions of quantum channels}
\label{Ap1}
In this appendix, we review different symmetry conditions for the density matrix and examine them in the decoherence models, particularly those involving only phase and qubit errors. This analysis is conducted from the perspectives of mixed states and Kraus operators, similar to analyzing the symmetry of pure states from state and Hamiltonian perspectives, respectively.

\begin{itemize}
    \item \textbf{Average (weak) symmetry condition:} Consider a state \(\ket{\Psi}\) in Hilbert space with a symmetry group \( G \). When a group element \( g \) acts unitarily on \(\ket{\Psi}\), the transformation is \(\ket{\Psi} \rightarrow U(g)\ket{\Psi}\). For a density matrix \(\rho\), a natural assumption is the action of \( U(g) \) and its adjoint \( U^\dagger(g) \) on \(\rho\), leading to 
    \begin{equation}
    \rho = U(g)\rho U^\dagger(g).
    \end{equation}
    This is known as the average or weak symmetry condition as it holds true only after taking the ensemble average in general. We note that a weak symmetry alone is not expected to support any SPT state because the sum of cohomology classes corresponding to \( U(g) \) and \( U^\dagger(g) \) is always zero.

    \item \textbf{Exact (strong) symmetry condition:} To define SPT phases in open quantum systems, a stronger symmetry condition is required. Specifically, one can consider the left and right operations separately, represented as 
    \begin{gather}\label{exsym}
    \rho = U(g_L)\rho U^\dagger(g_R),
    \end{gather}
    where \( g_{L,R} \in G \). We note that, for a pure state density matrix \(\rho = \ket{\Psi}\bra{\Psi}\), Eq.~\eqref{exsym} is nothing but the usual symmetry condition for a state vector. 
\end{itemize}

To discuss how different symmetry conditions apply to quantum channels given by completely positive trace preserving (CPTP) maps with symmetry \( G \), we consider the action of Kraus operators \( K_{\alpha} \) on the pure state \( \rho = \ket{\Psi}\bra{\Psi} \), where the transformation of the state by a CPTP map is given by \( \rho \rightarrow \sum_{\alpha} K_{\alpha} \rho K_{\alpha}^{\dagger} \). 

\begin{itemize}
    \item \textbf{Average (weak) symmetry condition:} Here, the combined action of all Kraus operators should respect the symmetry of the original state. This implies that the resulting state after the CPTP map should remain invariant under the group \( G \). The condition can be expressed as:
    \begin{gather}
     U(g) \left( \sum_{\alpha} K_{\alpha} \rho K_{\alpha}^{\dagger} \right) U^{\dagger}(g) = \sum_{\alpha} K_{\alpha} \rho K_{\alpha}^{\dagger} \quad \forall g \in G. 
    \end{gather}
    \textcolor{blue}{In the basis of Karus operators, it can be shown that
    \begin{gather}
        U(g)K_\alpha=e^{\ii\theta_\alpha}K_\alpha U(g)
    \end{gather}
    where the phase factor can't be canceled by redefinition of $U(g)$.}

    \item \textbf{Exact (strong) symmetry condition:} In this case, each Kraus operator must individually respect the symmetry. Specifically, each of \( \{K_{\alpha}\} \) must commute with the symmetry operation for all \( g \in G \) as follows:
    \begin{gather}
    U(g) K_{\alpha} = K_{\alpha} U(g) \quad \forall \alpha, \forall g \in G 
    \end{gather}
    This ensures that every quantum trajectory in the quantum channel preserves the symmetry.
\end{itemize}

 Reference~\cite{degroot2021symmetry} has shown that the strong symmetry condition can protect SPT phases in open quantum systems, with classification given by \(\mathcal{H}^{d+1}(G, U(1))\), where $G$ is the full symmetry group. Moreover, Ref.~\cite{ChongWang2023Average} has proposed that a combination of the average (weak) and exact (strong) symmetries can still protect SPT phases, with cohomology classification given by \(\sum_{p=1}^{d+1}\mathcal{H}^{d+1-p}(G,\mathcal{H}^p(K,U(1)))\), where \( G \) is average and \( K \) is exact. We remark that the models studied in Refs.~\cite{lee2022symmetry, chen2023symmetry} exhibit SPT phases protected by such mixed average-exact symmetry, where phase error is acted on either of vertex or edge qubits in the 2D cluster state. 
 Meanwhile, in the setup discussed in Sec.~\ref{Sec4} of the present work,   bit-flip error is acting on all qubits and, consequently, the SPT phase is protected by the exact symmetry. For the sake of completeness, we shall discuss this point in more detail below.

\begin{itemize}
    \item \textbf{Bit-flip error}: The density matrix for a state under bit-flip errors is given by 
    \[ \rho_D = \rho_D^A \rho_D^B = \sum_{\text{Dw}} e^{-H_{\text{Ising}}/2} \rho_{\text{Dw}} \sum_{\text{Ga}} e^{-H_{\text{Ising Gauge}}/2} \rho_{\text{Ga}}, \]
    where \( h \) and \( t \) in Eq.~(\ref{isgauge}) are set to be zero. First, we consider the \(\mathbb{Z}_2^{(0)}\) generator, which transforms \(\rho_D\) as \(\rho_D \rightarrow \prod_{v\in A}\tau_v^x \rho_D\). The operation \(\prod_{v\in A}\tau_v^x \rho_{\text{Dw}}\) results in a domain wall configuration where all \(\tau^x_v\) are transformed to identity, and identities on vertices are transformed into \(\tau^x_v\). In the classical Ising model, this operation flips all Ising spins. When the magnetic field \( h \) is zero, the energy before and after the flip remains the same, leaving \(\sum_{\text{Dw}} e^{-H_{\text{Ising}}/2} \rho_{\text{Dw}}\) unchanged. As a result, the left group action alone does not change the state \(\rho_D\), since \(\rho_D^A \rho_D^B\) commutes with each other's right group action. Similarly, a \(\mathbb{Z}_2^{(1)}\) generator acting on the gauge configuration \(\prod_{e\in\text{loop on }B}\sigma_e^x \rho_{\text{Ga}}\) only shifts the \(\mathbb{Z}_2\) flux in the corresponding Ising gauge theory from one place to another, resulting in a new state with the same energy. This fact can be easily verified from the Kraus operators' perspective, as every local quantum channel can be decomposed into \( K_i^1 = \sqrt{1-p_x}\mathbb{I}_i \) and \( K_i^2 = \sqrt{p_x}\sigma^x_i/\tau^x_i \), all of which should commute with the generators containing only Pauli \( X \) operators respectively.

   \item \textbf{Phase error on sub-lattice A (vertex qubits)}: With phase errors acting only on sub-lattice $A$, the density matrix is given by 
    \[ \rho_D = \rho_D^A \rho_D^B = \sum_{\text{Dw}} e^{-h\sum_i s_i}\rho_{\text{Dw}} \sum_{\text{Ga}}  \rho_{\text{Ga}}. \]
    When the $\mathbb{Z}_2^{(0)}$ generator acts on $\rho_{\text{Dw}}$ from either left or right, it will change $\sum_i s_i$ and result into a different state, i.e., $\prod_{v\in A}\tau_v^x  \sum_{\text{Dw}} e^{-h\sum_i s_i}\rho_{\text{Dw}}\neq \sum_{\text{Dw}} e^{-h\sum_i s_i}\rho_{\text{Dw}}$. To keep this term invariant, we need to act the symmetry generator from both left and right,  i.e., $\prod_{v\in A}\tau_v^x  \sum_{\text{Dw}} e^{-h\sum_i s_i}\rho_{\text{Dw}}\prod_{v\in A}\tau_v^x = \sum_{\text{Dw}} e^{-h\sum_i s_i}\rho_{\text{Dw}}$. Thus, the zero-form symmetry is average, while, from the similar discussion in the case of bit-flip error above, one can check that the one-form symmetry is exact. The resulting SPT phase is protected by  $\mathbb{Z}_{2,\text{avg}}^{(0)}\otimes\mathbb{Z}_2^{(1)}$.
 \item \textbf{Phase error on sub-lattice B (edge qubits)}: With phase errors acting only on sub-lattice $B$, the density matrix is given by 
    \[ \rho_D = \rho_D^A \rho_D^B = \sum_{\text{Dw}} \rho_{\text{Dw}} \sum_{\text{Ga}} e^{-t\sum_i s_i} \rho_{\text{Ga}}. \]
    It is evident that \(\mathbb{Z}_2^{(0)}\) remains invariant under both left and right actions. Meanwhile, the \(\mathbb{Z}_2^{(1)}\) symmetry generator flips \( s_i \) in the corresponding classical Ising model and changes the coefficient \( e^{-t\sum_i s_i} \) before \(\rho_{\text{Ga}}\). To restore the symmetry, we need to act the one-form symmetry generator on both the right and left sides simultaneously, rendering the one-form symmetry average. The resulting SPT phase is protected by $\mathbb{Z}_{2}^{(0)}\otimes\mathbb{Z}_{2,\text{avg}}^{(1)}$.
\end{itemize}

\bibliography{main}
\end{document}